\documentstyle[prd,aps,preprint]{revtex}
\newfont\fiverm{cmr5}
\input prepictex
\input pictex
\input postpictex

\begin{document}

\newcommand{\TeV}{\,{\rm TeV}}
\newcommand{\GeV}{\,{\rm GeV}}
\newcommand{\MeV}{\,{\rm MeV}}
\newcommand{\keV}{\,{\rm keV}}
\newcommand{\eV}{\,{\rm eV}}
\def\ap{\approx}
\def\beqar{\begin{eqnarray}}
\def\eeqar{\end{eqnarray}}
\newcommand{\bea}{\begin{eqnarray}}
\newcommand{\eea}{\end{eqnarray}}
\def\ler{\lesssim}
\def\gtr{\gtrsim}
\def\beq{\begin{equation}}
\def\eeq{\end{equation}}
\def\haf{\frac{1}{2}}
\def\plb#1#2#3#4{#1, Phys. Lett. {\bf #2B} (#4) #3}
\def\plbb#1#2#3#4{#1 Phys. Lett. {\bf #2B} (#4) #3}
\def\npb#1#2#3#4{#1, Nucl. Phys. {\bf B#2} (#4) #3}
\def\prd#1#2#3#4{#1, Phys. Rev. {\bf D#2} (#4) #3}
\def\prl#1#2#3#4{#1, Phys. Rev. Lett. {\bf #2} (#4) #3}
\def\mpl#1#2#3#4{#1, Mod. Phys. Lett. {\bf A#2} (#4) #3}
\def\rep#1#2#3#4{#1, Phys. Rep. {\bf #2} (#4) #3}
\def\lpp{\lambda''}
\def\ccg{\cal G}
\def\slash#1{#1\!\!\!\!\!/}
\def\rpv{\slash{R_p}}

\setcounter{page}{1}
\preprint{\begin{tabular}{r} KAIST-TH 99/10 \\
SCIPP-99/51  \\
hep-ph/9912218 \end{tabular}}

\title{Quintessence, Flat Potential  and String/$M$ Theory Axion}

\author{Kiwoon Choi}

\medskip

\address{
Santa Cruz Institute for Particle Physics, Santa Cruz,
CA 95064, USA \&
\\
Department of Physics, Korea Advanced Institute of Science and Technology\\
        Taejon 305-701, Korea}

\tighten
\maketitle

\begin{abstract}
A slow-rolling scalar field (Quintessence)  has been proposed
as the origin of accelerating universe at present.
We discuss some  features of quintessence,  particularly those
related with the {\it flatness} of its potential.
We distinguish two types of quintessences, a metrical
quintessence which couples to the  particle physics sector 
{\it only} through the mixing  with the spacetime metric
and a non-metrical one 
without such property.
It is stressed that these two types of quintessences have quite different
features in regard to the  fine tuning of parameters required for
the flat potential, as well as different implications
for the  quintessence-mediated long range  force.
The fine tuning argument strongly suggests that    non-metrical quintessence  
should correspond to the Goldstone boson of an almost exact nonlinear 
global symmetry whose explicit breaking by quantum gravity effects
is highly suppressed.
It  is briefly  discussed   whether the radion field in higher
dimensional theories can be a candidate for
metrical quintessence.
We finally  discuss the possibility that   string/$M$ theory
axion corresponds to a Goldstone-type quintessence,
and argue that a certain  combination of the
heterotic $M$ theory or Type I string theory axions
can be a good candidate for quintessence
if some conditions on the moduli dynamics
are satisfied.

\end{abstract}



\section{introduction}

Recent measurements of the luminosity-red shift relation for Type Ia supernovae
suggest that the Universe is accelerating and thus
a large fraction of the energy density   has negative
pressure \cite{accel}. In fact,  even before these supernovae data, 
there have been some
indications that the Universe is dominated by  an yet unknown form of 
smooth dark energy density \cite{turner}.
The Big-Bang nucleosynthesis implies the baryon mass density
$\Omega_B\sim 0.05$, while  the measurements  of cluster masses suggest
the (clumped) dark matter mass density  $\Omega_{DM}\sim 0.3$.
(Here all mass  densities are expressed  in terms of their ratios
to the critical mass density $\rho_c=3M_P^2H^2$.)
On the other hand, the anisotropy of the cosmic microwave background 
shows the first peak
of its angular power spectra at $l\sim 200$, implying that  the total
energy density $\Omega_{\rm TOT}\sim 1$
which is consistent   with the prediction of inflation.
This suggests 
the existence of a smooth (unclumped) dark
energy density with  $\Omega_X\sim 0.7$,
and  the supernovae data  provide a strong support  for such dark
energy density with further information
that it has  negative pressure density $p_X\lesssim -0.6\Omega_X$.

If one accepts the accelerating universe as a reality, the utmost question is
what is the nature of the dark  energy with negative pressure. 
Of course, a nonvanishing cosmological constant, i.e. a non-evolving
vacuum energy density, 
would be the simplest form of dark energy density with negative pressure.
However there has been some  prejudice that the correct, if  exists any, 
solution to the cosmological constant problem \cite{weinberg} will lead to 
an exactly vanishing vacuum energy density.
Recently quintessence in the form of  slow-rolling scalar field 
has been proposed as an alternative form of dark energy density 
with negative pressure \cite{quint,quint1,quint2}.
The true minimum of the quintessence potential is presumed to vanish.
However the present value of $Q$ is {\it displaced} from the true minimum,
providing 
\beq
V_Q\sim H_0^2M_P^2\sim (3\times 10^{-3} \, {\rm eV})^4,
\label{quintpotential}
\eeq
and a negative pressure with the equation of state:
\beq
\omega=\frac{p_Q}{\rho_Q}=\frac{\frac{1}{2}{\dot{Q}}^2-V_Q}{
\frac{1}{2}{\dot{Q}}^2+V_Q} \ler -0.6,
\eeq
where $M_P=1/\sqrt{8\pi G_N}=2.4\times 10^{18}$ GeV denotes the
reduced Planck scale and $H_0\sim 10^{-32}$ eV is the Hubble
expansion rate at present. 
In fact, when combined with the equation of motion for $Q$, 
the negative pressure value  requires the following slow roll
condition:
\beq
\left|\frac{\partial V_Q}{\partial Q}\right|\lesssim
\left|\frac{V_Q}{M_P}\right|.
\label{slowroll}
\eeq
This condition indicates that if $V_Q$
is well approximated by the  terms with nontrivial $Q$-dependence,
not by a $Q$-independent constant term,  
the typical range of $Q$ is of the order of
$M_P$ or at least not far below $M_P$.

The basic idea of quintessence is to assume a scalar field
whose potential energy $V_Q$ is  within  
the order of $H_0^2M_P^2$ over the Planck scale range of $Q$.
Then an utmost question is how such an extremely flat $V_Q$
can arise from  realistic particle physics models \cite{lyth,choi0}.
In section II, we discuss this issue together with
some generic  features of quintessence.
We distinguish two types of quintessences, a Brans-Dicke type
metrical quintessence
which couples to the particle physics sector only through the mixing
with the conformal factor of the spacetime metric
and a non-metrical one without such property.
These two types of quintessences
have quite different  features in regard to the fine
tuning of parameters required for the   flat potential,
as well as different implications for the quintessence-mediated 
long range force. 
We argue that non-metrical  quintessence should corrrespond
to the Goldstone boson of an almost exact nonlinear global symmetry
whose explicit breaking by quantum gravity effects is highly suppressed.
Unless  a Goldstone-type, non-metrical quintessence 
encounters a severe fine tuning problem which is much worse than
the conventional cosmological constant problem.
To make this point more clear, we briefly discuss the difficulties
of a non-Goldstone type  non-metrical quintessence having
an inverse-power potential $V_Q=m^{4+\alpha}/Q^{\alpha}$.
We also discuss whether the radion field in theories
with (large) extra dimensions \cite{savas}
 can be a candidate for metrical quintessence \cite{banks0}.
In section III, we note  that 
a certain combination of axions in
heterotic $M$-theory \cite{horava}
or  Type I string theory \cite{munoz,polchinski}
can be a  plausible candidate for non-metrical
quintessence \cite{choi0} if its  modulus partner is stabilized 
at  a large value through  the K\"ahler stabilization
mechanism \cite{banks1}.
Although this quintessence axion  could avoid  
the fine tunings of parameters associated with its flat potential,
it still suffers from the fine tuning
of cosmological initial conditions.
In section IV, we   discuss a late time inflation scenario
based on the modular and CP invariance of the effective potential,
which would solve this   initial condition problem.

\section{some generic features of quintessence}

The supernovae data indicate that the Universe has been accelerating
since the redshift factor $z={\cal O}(1)$,
and thus over the time scale of  $t_0\sim 1/H_0\sim 10^{10}$ yrs
\cite{accel}.
To have a negative pressure $p_Q=\dot{Q}^2/2-V_Q$, one needs
\beq
\dot{Q}\lesssim \sqrt{V_Q}\sim H_0 M_P
\label{slowroll2}
\eeq
over the accelerating period.
In order for the above condition to be satisfied, one needs also 
\beq
\ddot{Q}\lesssim H_0^2 M_P
\label{slowroll3}
\eeq
over the same accelerating period.
Note that we have ignored the coefficients of order one,
and also $\dot{Q}$ and
$\ddot{Q}$ can be regarded  as appropriately averaged quantities. 
At any rate, when combined with the equation of motion
\beq
\ddot{Q}+3H\dot{Q}+\frac{\partial V_Q}{\partial Q}=0,
\eeq
the Eqs.(\ref{slowroll2}) and (\ref{slowroll3}) lead to
\beq
\left|\frac{\partial V_Q}{\partial Q}\right|\ler 
\left|\frac{V_Q}{M_P}\right|\sim \frac{(3\times 10^{-3} \, {\rm eV})^4}{M_P},
\label{slowroll1}
\eeq
which has to be satisfied over the range of $Q$ from the value at
$z\sim 1$ to the present value.

To see the implications of the slow-roll condition (\ref{slowroll1}),
let us assume that $V_Q$ is well approximated by 
\beq
V_Q=V_0+\frac{1}{2}m_Q^2 Q^2,
\eeq
where $V_0$ is a $Q$-independent constant.
If $V_Q$ is dominated by the constant $V_0$ during the period 
of interest, Eq.(\ref{slowroll1}) leads to
\beq
|Q/M_P|\lesssim (H_0/m_Q)^2\sim
(10^{-32} \, {\rm eV}/m_Q)^2
\label{qbound}
\eeq
and thus
\beq
|m_Q^2 Q^2/V_0|\lesssim  H_0^2/m_Q^2\sim (10^{-32} \, {\rm eV}/m_Q)^2.
\eeq
This shows that if $m_Q\gg H_0\sim 10^{-32}$ eV,
we are in the limit of {\it non-evolving} cosmological constant in which
$Q$ is frozen at the minimum of $V_Q$.
(Here  $m_Q^2$ is assumed  positive.)

However if the $Q$-dependent part $m_Q^2 Q^2/2$ dominates or at least
non-negligible compared to the constant $V_0$, 
Eq.(\ref{slowroll1}) leads to
\beq
|Q|\gtrsim M_P, \quad
m_Q\lesssim 10^{-32} \, {\rm eV},
\eeq
and so {\it $V_Q$ must be extremely flat over the Planck scale range of $Q$}.
This is true even when $m_Q^2Q^2/2$ 
is replaced by other forms of potential, e.g. the inverse power potential
$m^{4+\alpha}/Q^{\alpha}$ or the exponential potential $m^4 e^{-Q/v}$.
Thus if it has any feature  distinguished from
the non-evolving cosmological constant, 
the quintessence model  involves an extremely flat  potential,
$V_Q\sim H_0^2M_P^2\sim (3\times 10^{-3} \, {\rm eV})^4$
over the  Planck scale range of $Q$,  and so   an almost massless
scalar boson  with  mass 
$m_Q\sim H_0\sim 10^{-32}$ eV in rough order of magnitude estimate.

As is well known, there are rather strong observational limits
on the couplings (and also their evolutions) between an almost massless
scalar and  normal matters \cite{damour}.
For a discussion of these constraints and also of
the fine tunings required for
the flat potential, it is convenient to distinguish
two types of quintessences: a 
metrical quintessence which couples
to the particle physics sector only through the mixing with the spacetime
metric, and a non-metrical one without such feature.
Note that metrical quintessence is quite similar to the Brans-Dicke
type scalar field in scalar-tensor gravity models.
As we will see, these two types of quintessences
have quite different features in
the fine tuning of parameters for the flat potential,  as well as
its implications for the quintessence-mediated long range force.

Let us first consider non-metrical quintessence $Q$
whose couplings to matter are {\it not} related with those of
the spacetime metric.
The  interaction lagrangian  can  include
\beq
{\cal L}_{int}=\frac{Q}{M_P}
(\xi_{F^2}F_{\mu\nu}F^{\mu\nu}
+\xi_{G^2}G^a_{\mu\nu}G^{a\mu\nu}
+\sum_q \xi_q m_q\bar{q}q),
\label{quintcoupling}
\eeq
where  $F_{\mu\nu}$ and
$G^a_{\mu\nu}$ denote the electromagnetic and gluon field
strengths, respectively, and $q$ stands for the light quarks with mass
$m_q$.
Note that $\xi_I$'s represent the ratios of the
quintessence couplings  to the gravitational
coupling. 
Since the operator combination which couples to $Q$
does not  match to  the energy momentum tensor even in the 
non-relativistic limit, they violate the equivalence principle
for non-relativistic laboratory bodies. 
More explicitly, the relevant  operator combination can be written as
\beq
\xi_{G^2}G^{a\mu\nu}G^a_{\mu\nu}+\sum_q\xi_qm_q\bar{q}q=
\frac{2g}{\beta(g)}\xi_{G^2}T^{\mu}_{\mu}
+\sum_q [\xi_q-\frac{2g}{\beta(g)}\xi_{G^2}(1+\gamma_m)]m_q\bar{q}q,
\eeq
where $T^{\mu}_{\mu}$ is  the trace of the QCD energy 
momentum tensor with the QCD beta function
$\beta (g)$ and the mass anomalous dimension $\gamma_m$.
Then the second term in the r.h.s. gives rise to a
composition-dependent acceleration $a_i$ of the $i$-th test body, yielding
\beq
\frac{a_i-a_j}{a_i+a_j}
\approx  \left(
\frac{2g}{\beta(g)}\xi_{G^2}+\frac{\langle \sum_q\tilde{\xi}_q
m_q\bar{q}q\rangle_s}{\langle   T^{\mu}_{\mu}\rangle_s}\right)
\left(\frac{\langle \sum_q\tilde{\xi}_qm_q\bar{q}q\rangle_i}{\langle
T^{\mu}_{\mu}\rangle_i}-
\frac{\langle \sum_q\tilde{\xi}_qm_q\bar{q}q\rangle_j}{
\langle T^{\mu}_{\mu}\rangle_j}\right)
\eeq
where $\tilde{\xi}_q=\xi_q-2g\xi_{G^2}(1+\gamma_m)/\beta(g)$
and the subscript ``$s$'' denotes the source of the
long range force.
Under the assumption that there is no sizable cancellation between
$\xi_q$ and $2g\xi_{G^2}(1+\gamma_m)/\beta(g)$, which is the case
for  non-metrical quintessence, the current
observational limit  on  non-universal acceleration \cite{damour}   
\beq
|\Delta a/a|\lesssim 10^{-12}
\eeq
leads to \cite{carroll}
\beq
\xi_{G^2}\lesssim {\cal O}(10^{-5}),
\quad
\xi_q\lesssim {\cal O}(10^{-5}).
\label{oblimit}
\eeq
for the operator coefficients $\xi_{G^2}$ and $\xi_q$ renormalized
at energy scale $\sim 1$ GeV.

If $\xi_{F^2}$ is nonzero at low energies, say at ${\cal O}(1)$ eV,
a time-varying $Q$ leads  to a time-varying  fine structure constant
which is constrained \cite{damour} as 
\beq
 \dot{\alpha}_{em}/{\alpha_{em}} \lesssim 10^{-16} \, {\rm yr}^{-1}.
\label{wilco}
\eeq
Obviously this gives
\beq
\xi_{F^2}\dot{Q}\lesssim 3\times 10^{-6} M_PH_0,
\eeq
a rather strong observational limit on the combination
$\xi_{F^2}\dot{Q}$.

The direct observational limits (\ref{oblimit}) already suggest
that the non-derivative couplings between non-metrical
$Q$ and  normal matter
are highly suppressed, even compared to the gravitational
strength. 
In fact, although  indirect, much stronger limits on generic 
non-derivative couplings
can be obtained from   the  fine tuning argument associated with
the extremely flat  potential.
The quintessence couplings of Eq. (\ref{quintcoupling})
mean that the electromagnetic and QCD fine structure constants,
and the light quark masses are all $Q$-dependent.
Then the QCD and electroweak physics contribute to $V_Q$,
yielding  the following
form of  the quintessence potential 
in rough order of magnitude estimates
\beq
V_Q=\Lambda_{QCD}^4+m_q\Lambda_{QCD}^3+\frac{1}{16\pi^2}
M_W^4\ln (M_W^2/\mu^2)+...
\label{nonmetricalpotential}
\eeq
where  the ellipsis stands for  other types of contributions
to $V_Q$.
Here  the QCD scale $\Lambda_{QCD}\propto e^{-2\pi/b\alpha_{QCD}(Q)}$
and the W-mass $M_W\propto \alpha_{em}(Q)$ are some functions
of $Q$ whose $Q$-dependence would be determined by  the $Q$
couplings to the gluon and electroweak field strengths.

Suppose we tuned the parameters of the theory to make
$V_Q\sim (3\times 10^{-3} \, {\rm eV})^4$ for {\it a particular
value of $Q$}, say for $Q=Q_0$. Then for different value of $Q$,
e.g. $Q=Q_0+\Delta Q$, the three terms in $V_Q$  of 
Eq.(\ref{nonmetricalpotential}) would lead to a potential
energy difference 
\beqar
&& \delta_1 V_Q\sim \xi_{G^2}\Lambda_{QCD}^4\frac{\Delta Q}{M_P}
\sim 10^{43} \, \xi_{G^2}H_0^2M_P^2\frac{\Delta Q}{M_P},
\nonumber \\
&&\delta_2 V_Q\sim \xi_q m_q\Lambda_{QCD}^3\frac{\Delta Q}{M_P}
\sim 10^{42} \, \xi_q H_0^2M_P^2\frac{\Delta  Q}{M_P},
\nonumber \\
&&\delta_3 V_Q\sim \
10^{-2}\xi_{F^2}M_W^4\frac{\Delta Q}{M_P} \sim 10^{52} \, \xi_{F^2}
H_0^2M_P^2 \frac{\Delta Q}{M_P}.
\label{variation}
\eeqar
These  estimates of the variation of $V_Q$  show that
even after the fine tunining for
$V_Q\sim H_0^2M_P^2$ at $Q=Q_0$, one still needs 
additional fine tunings in order to keep
$V_Q$ within the order of
$H_0^2M_P^2$ 
over the  Planck scale range of $Q$. 
Generically, one  needs {\it different}  fine tunings
 for {\it different} values of $Q$, and then
{\it the theory would  suffer from 
the infinite numbers of fine tuning problems}.

There are essentially two ways to avoid this difficulty.
The first one is to assume a non-linear continuous global symmetry
($U(1)_Q$)
under which $Q$ is shifted by a constant.
Such $U(1)_Q$   would assure that
all non-derivative couplings of $Q$ to the particle physics sector
are suppressed, so that the particle physics sector 
contribution to  $V_Q$ can be small enough for the wide range of $Q$. 
The second possibility is that although
the non-derivative couplings between $Q$ and normal matter are sizable, 
they are all precisely
correlated to each other 
in such a way that their  contributions
to $V_Q$  are cancelled {\it independently of}
the  values of $Q$.
It is likely that the only sensible model in this direction
is  the metrical quintessence whose couplings to matter are induced
{\it only} through  the mixing with the spacetime metric.
This leads us to  conclude that, in order to avoid the infinite
numbers of fine tuning problems, non-metrical quintessence 
should correspond to the Goldstone boson of a nonlinear global
symmetry under which
\beq
U_Q(1): 
\quad Q\rightarrow Q+{\rm constant}.
\eeq

Still the potential difficulty for Goldstone-type quintessence
is that it  requires a nonzero but extremely tiny breaking
of $U(1)_Q$.
Note that generically a  global symmetry can not be arbitrarily good,
particularly when quantum gravity effects are included.
So the questions are
(1) how $U(1)_Q$ could avoid a potentially large breaking from
quantum gravity effects
and (2) what is the origin of the tiny $U(1)_Q$ breaking
which would provide  small but nonvanishing  $V_Q\sim H_0^2M_P^2$.
To quantify this issue,
let us first estimate the allowed size of $U(1)_Q$ breaking.
Again assuming that $U(1)_Q$ breaking couplings of
Eq.(\ref{quintcoupling}) are  not correlated to each other, 
the variations of
$V_Q$ in Eq.(\ref{variation}) suggest that
\beq
\xi_{F^2}\lesssim 10^{-52},
\quad
\xi_{G^2}\lesssim 10^{-43},
\quad
\xi_q\lesssim 10^{-42},
\label{limit}
\eeq
in order for $V_Q$ remain to be within the order of
$H_0^2M_P^2$
over the Planck scale range of $Q$.
This already shows   that $U(1)_Q$-breaking  couplings
should be extremely  suppressed, even compared to the gravitational
coupling  strength.

In fact, the variation (\ref{variation}) of $V_Q$
includes  only the contributions from the particle physics 
degrees of freedom
with momenta $p\lesssim M_W$. If one includes the contributions
from  higher momentum modes, the typical variation of $V_Q$
takes much larger value.
As an illustrative example, let us consider
generic  supergravity models with
cutoff scale $\Lambda\sim M_P$. The K\"ahler potential $K$,
superpotential $W$, and gauge kinetic functions $f_a$ can be expanded
in powers of the visible sector matter fields $\Phi^i$:
\beqar
&&K=K_0(Z,Z^*)+Z_{ij}(Z,Z^*)\Phi^i\Phi^{*j}
+H_{ij}(Z,Z^*)\Phi^i\Phi^j+
X_{ijk}(Z,Z^*)\Phi^i\Phi^j\Phi^{*k}+...,
\nonumber \\
&&  W=W_0(Z)+Y_{ijk}(Z)\Phi^i\Phi^j\Phi^k+
\Gamma_{ijkl}(Z)\Phi^i\Phi^j\Phi^k\Phi^l+...,
\nonumber \\
&& f_a=f_{a0}(Z)+f_{aij}(Z)\Phi^i\Phi^j+...,
\eeqar
where $Z$ denotes the  chiral multiplet  including the
quintessence degree of freedom $Q$.
Any $Q$-dependence of the coefficient functions
in the expansion gives  rise to non-derivative couplings
between $Q$ and $\Phi^i$,  so  reflects the explicit  breaking of
$U(1)_Q$.
The typical sizes of the variation of $V_Q$  induced by 
each of the $Q$-dependent coefficient  functions 
are estimated in \cite{choi0}, taking into account
various types of quantum corrections.  Again if any  of 
the non-derivative couplings is  sizable, 
we can not keep $V_Q$ small enough over the
Planck scale range of $Q$ by tuning finite number of parameters. 
This consideration leads to the following
strong limits on the  $U(1)_Q$ breaking couplings
(in the unit with $M_P=1$):
\beqar
&& \frac{\partial}{\partial Q}\ln (K_0)\lesssim 10^{-86},
\quad \frac{\partial}{\partial Q}\ln (W_0)\lesssim 10^{-86},
\nonumber \\
&&  \frac{\partial}{\partial Q}Z_{ij}\lesssim 10^{-81},
\quad \frac{\partial}{\partial Q}\left|\frac{\partial H_{ij}}
{\partial Z^*}\right|^2\lesssim 10^{-74},
\nonumber \\
&&  \frac{\partial}{\partial Q}\left|X_{ijk}\right|^2\lesssim
10^{-74}, \quad
\frac{\partial}{\partial Q}\left|Y_{ijk}\right|^2\lesssim
10^{-79},
\nonumber \\
&& \frac{\partial}{\partial Q}\left|\Gamma_{ijkl}\right|^2
\lesssim 10^{-72}, \quad
\frac{\partial}{\partial Q}\left|\frac{\partial f_{aij}}
{\partial Z}\right|^2\lesssim 10^{-67},
\nonumber \\
&& \frac{\partial}{\partial Q}{\rm Re}(f_{a0})
\lesssim 10^{-81},
\quad
\frac{\partial}{\partial Q}{\rm Im}(f_{QCD})
\lesssim 10^{-42},
\label{bound}
\eeqar
where we have assumed an weak scale gravitino mass
and   the last bound is on the coupling of $Q$ to the QCD anomaly
$G^{a\mu\nu}\tilde{G}^a_{\mu\nu}$.
In fact, even the flat potential argument does  not  give 
a  meaningful limit on the coupling to the electroweak
anomaly. So  it is in principle possible that $\xi_{F\tilde{F}}$
is sizable for the coupling
\beq
\xi_{F\tilde{F}}\frac{Q}{M_P}F^{\mu\nu}\tilde{F}_{\mu\nu}.
\eeq
Unless $|\xi_{F\tilde{F}}|\ll 10^{-2}$,
such coupling may lead to an observable rotation of the polarization
axis of the radiation from remote sources
\cite{carroll,sikivie}.
However in view of the gauge coupling unification,
it  is somewhat unlikely that $Q$ has a   
sizable coupling to $F\tilde{F}$, while its coupling to
$G\tilde{G}$ is so small as given by the last bound of
Eq.(\ref{bound}).

The above limits on the $U(1)_Q$ breaking couplings 
suggest that $U(1)_Q$ must be an almost exact
global symmetry.
It is highly nontrivial to have such an almost exact
global symmetry in any realistic models for particle physics
once the effects of Planck scale physics are included through
arbitrary higher dimensional operators in the effective action.
In the next  section, we will explore the possibility of such
global symmetry in the framework of string/$M$ theory
and point out that a certain  combination of the heterotic
$M$ theory or Type I string theory axions
can be a plausible candidate for quintessence if some 
conditions on the moduli dynamics are satisfied.
Here we stress that $U(1)_Q$ has nothing to do with
a small $V_Q$ at its minimum.
It just ensures that $V_Q$
is  small over the Planck scale  range of $Q$ once it were small
at its minimum.
Without $U(1)_Q$,  non-metrical quintessence model
would suffer from infinite numbers of fine tuning problems
to keep $V_Q$ to be small  over the wide range of $Q$. 
At any rate, the limits (\ref{limit})   which are based on the
fine tuning argument suggest that it is  hopeless to see
the long range force mediated by non-metrical quintessence.

Let us now discuss the Brans-Dicke type metrical quintessence which couples
to matter only through the mixing with the conformal  factor
of the spacetime metric.
We first consider the observational limits
on its couplings.
The dynamics of  metrical quintessence would be described by the following
effective  action:
\beq
S_{eff}=\int d^4 x \left[\sqrt{-g}\left(-\frac{1}{2}M_P^2{\cal R}(g)
+\frac{1}{2}g^{\mu\nu}\partial_{\mu}Q\partial_{\nu}Q-V_0(Q)\right)
+\sqrt{-\tilde{g}} \, {\cal L}_{m}(\tilde{g}_{\mu\nu},\Phi)\right]
\eeq
where ${\cal R}$ is the Ricci scalar density
and  ${\cal L}_m$ denotes  the lagrangian density of
generic matter fields
$\Phi$ (including the conventional particle physics sector)
under the background spacetime metric given by 
\beq
\tilde{g}_{\mu\nu}=A^2(Q/M_P)  g_{\mu\nu}.
\eeq
Then the  couplings between $Q$ and $\Phi$ are given by 
\beq
{\cal L}_{int}=Q\frac{\partial}{\partial Q}\left(
\sqrt{-\tilde{g}}{\cal L}_{m}\right)_{Q=Q_0}=\xi
\,  \frac{Q}{M_P}T^{\mu}_{\mu}
\eeq
where
\beq
\xi=M_P\left(\frac{\partial \ln A^2}{\partial Q}\right)_{Q=Q_0}
\eeq
and 
\beq
T^{\mu}_{\mu}= \frac{\beta(g)}{2g}G^{a\mu\nu}G^a_{\mu\nu}
+\frac{\beta(e)}{2e}F^{\mu\nu}F_{\mu\nu}+ \sum_i (1+\gamma_m)m_i
\bar{\psi}_i\psi_i+...
\label{energymomentum}
\eeq
denotes the trace of the energy momentum tensor
of the matter fields $\Phi=G^a_{\mu\nu}, F_{\mu\nu}, \psi_i.$
Since it respects the equivalence principle in non-relativistic limit,
the above quintessence couplings   lead to  a deviation
from the Einstein's general relativity only through the relativistic
corrections.    
Currently performed gravitational experiments
in the solar system provide upper bounds on
such deviation \cite{damour}, implying 
\beq
\xi\lesssim 4.5\times 10^{-2}.
\label{limit1}
\eeq
One might think   that the strong observational limit
(\ref{wilco})  on
the time-varying  fine structure  constant can give
a  constraint on the evolution of metrical quintessence.
However existing  limits are derived from
the electromagnetic processes with momentum transfer 
much smaller than the electron mass. For such low momentum
transfer, we have $\beta(e)=0$ and so a time-varying metrical
quintessence does {\it not} lead to a time-varying
electromagnetic fine structure constant.

Let us now turn to the flatness of the metrical quintessence
potential.
Since the couplings between  $Q$ and the particle physics sector
are determined by the mixing with the spacetime metric,
the whole contribution to $V_Q$ from the particle physics sector
is simply given by
\beq
\Delta S_{eff}\equiv \int d^4x \sqrt{-g} \, \Delta V_Q
=\int d^4x \sqrt{-\tilde{g}}\Lambda_m^4=
\int d^4 \sqrt{-g}A^4(Q/M_P)\Lambda_m^4,
\eeq
where $\Lambda_m^4$ is a {\it $Q$-independent} energy density
which is  a generic function of all
particle physics parameters in ${\cal L}_m$.
Note that the general covariance w.r.t $\tilde{g}_{\mu\nu}$
guarantees the above form of $\Delta V_Q$ which is obtained
after integrating out all particle physics sector fields. 
The  total quintessence  potential energy 
is then given by
\beq
V_Q=V_0+\Delta V_Q =V_0(Q/M_P)+A^4(Q/M_P)\Lambda_{m}^4.
\eeq

An interesting  feature of the  metrical quintessence potential
is that  once the particle physics parameters
are adjusted to yield $\Lambda^4_m\lesssim H_0^2M_P^2$,
the entire contribution to $V_Q$ from the particle physics sector
remains to be within the order of $H_0^2M_P^2$
over  the Planck scale range of $Q$.
Of course one still needs to adjust the theory to keep
$V_0$ to be small enough. The typical mass scales of the physics 
responsible
for $V_0$  can be much lower than those of the conventional
particle physics
and/or the physics generating $V_0$ may enjoy extra symmetry
which would assure that $V_0$ is small enough.
In this case, one may be able to make the total $V_Q=V_0+\Delta V_Q$
flat enough without a severe fine tuning problem other than
the  minimal fine tuning  for small $\Lambda_m^4$.
Note that if  the (non-derivative) coupling between metrical 
quintessence and  normal matter takes a value not far below
the current observational limit (\ref{limit1}),
e.g. $\xi={\cal O}(10^{-2})$,
it can lead to observable effects
in  future gravity experiments.

An interesting
candidate for metrical quintessence
 would be  the radion field in theories with 
large extra dimensions \cite{savas,banks0}.
However at least in the simplest framework,
the  radion has  too  strong  coupling with  matter to be
a quintessence. 
For example, if one considers a higher dimensional theory
with $n$  large flat dimensions with  common radius,
the canonical radion field $Q$ in the Einstein frame of the
four dimensional metric $g_{\mu\nu}$ appears in
the $(4+n)$-dimensional metric:
\beq
ds^2_{(4+n)}=g^{(4+n)}_{MN}dx^Mdx^N=
e^{\xi Q/M_P}g_{\mu\nu}dx^{\mu}dx^{\nu}+
 e^{-2\xi Q/nM_P}dy^idy^i
\eeq
where
\beq
\xi=\sqrt{2n/(n+2)}.
\eeq
If the whole particle physics sector lives on a three brane,
the radion couples to the particle physics sector
only through the induced  metric on the brane:
\beq
g^{(4+n)}_{\mu\nu}=e^{\xi Q/M_P} g_{\mu\nu}.
\eeq
Obvioulsy, the value of $\xi$ in this framework
does not satisfy the observational limit (\ref{limit1}),
and thus   the  radion $Q$ can not be a quintessence.
(Of course, once one makes the radion massive by suitable stabilization
mechanism \cite{stabil},  the limit (\ref{limit1}) can be avoided.)
Although a Brans-Dicke type quintessence
can not arise from the simplest version,
it may be possible to  arrange  the higher dimensional theory
and the compactification scheme to yield an almost massless
radion field with $\xi\lesssim {\cal O}(10^{-2})$.

Let us close this section with a brief discussion of
the difficulties of  specific non-metrical quintessence which is
{\it not} Goldstone type, but has  an inverse power law potential: 
$V_Q\sim m^{4+\alpha}/Q^{\alpha}$.
Perhaps the most attractive feature of this model would be 
the  tracker  behavior
that $\rho_Q\sim \rho_{matter}$ at present for   wide range
of initial conditions of $Q$ and $\dot{Q}$.
This feature allows us to avoid the cosmic coincidence 
puzzle asking why now $\rho_Q$
is comparable to $\rho_{matter}$
although their  cosmological evolutions are generically  different
\cite{quint1,quint2}.

It has been pointed out  \cite{binetruy}
that the inverse power potential
can be generated by  nonperturbative dynamics in the underlying
supersymmetric theory \cite{ads}.
The model includes  a hidden supersymmetric QCD sector
with $SU(N_c)$ gauge group and
$N_f$ flavors  of hidden quarks and antiquarks  
$\Phi\equiv (\Phi_i,\Phi^c_i)$.
It is then assumed that the tree level superpotential
is {\it  independent} of $\Phi$.
The $\Phi$-dependent part of the exact  superpotential is
then given by 
\beq
W_{dyn}=\left(\frac{\Lambda^{3N_c-N_f}}{{\rm det}(\Phi_i\Phi_j^c)}
\right)^{1/(N_c-N_f)},
\label{ads}
\eeq
where $\Lambda$ denotes the dynamical scale of the hidden
$SU(N_c)$ gauge group.
If $W_{dyn}$ is a dominant term in the full superpotential,
the effective potential of the $SU(N_f)$ invariant $D$-flat direction
$\Phi_i=\Phi^c_i\equiv Q$
is given by
\beq
V_Q={\cal F}\Lambda^{4+\alpha}/{Q^{\alpha}},
\eeq
where $\alpha=2(N_c+N_f)/(N_c-N_f)$ and ${\cal F}$ is a function
of $Q/M_P$  which is essentially of order unity.
The explicit form of ${\cal F}$ depends on the
K\"ahler potential of $\Phi$.
For $\Lambda\sim   M_P(H_0/M_P)^{2/(4+\alpha)}\sim 10^{-120/(4+\alpha)}M_P$,
this potential is within the
range of $H_0^2M_P^2$  over the Planck scale range of $Q$,
so can be a candidate for quintessence potential.

However the problem is that $W_{dyn}\sim H_0M_P^2\ll m_{3/2}M_P^2$ for
$Q\sim M_P$, so it can {\it not} be a dominant
term in the superpotential in any realsitic model of supersymmetry breaking.
For $Q\sim M_P$, supergravity effects become important, and so
we have to consider $V_Q$ in the supergravity framework
in which the scalar potential is schematically given by
\beq
V_{sugra}=|F|^2+D^2-\frac{3|W|^2}{M_P^2},
\eeq
where $F$ and $D$ denote the auxiliary $F$ and $D$
components of supersymmetry breaking fields, and $W$ is the total
superpotential.
For realistic phenomenology, we need $|F|\gtrsim M_W^2$
(and/or $D\gtrsim M_W^2$)
for the weak scale mass $M_W$.
Obviously we then need   
$W\sim M_P|F|\sim m_{3/2}M_P^2$ 
in order to have the total vacuum energy density much smaller than $M_W^4$.
Since $W_{dyn}$ of Eq.(\ref{ads}) is  just  of the order of $H_0M_P^2$,
the total superpotential must include an additional term 
$W_0\sim m_{3/2}M_P^2$.
Once $W_0$ is introduced, which is a necessity, then
the flatness of $V_Q$ is totally lost.

To see this  explicitly, let us consider the case with
\beqar
W&=&W_0+W_{dyn}+\epsilon_1\frac{(\Phi_i\Phi_i^c)^n}{M_P^{2n-3}}+...,
\nonumber \\
K&=&\Phi_i\Phi_i^*+\Phi_i^c\Phi^{c*}+\epsilon_2 
\frac{(\Phi_i\Phi_i^*)^k}{M_P^{2k-2}}+...,
\eeqar
where $n$ and $k$ are some  integers which can be arbitrarily
large, and the ellipses
stand for other possible terms.
We then have (again schematically)
\beq
V_Q=\frac{\Lambda^{4+\alpha}}{Q^{\alpha}}+
m^2_{3/2}Q^2+\epsilon_2m_{3/2}^2\frac{Q^{2k}}{M_P^{2k-2}}
+\epsilon_1m_{3/2}\frac{Q^{2n}}{M_P^{2n-3}}
+\epsilon_1^2\frac{Q^{4n-2}}{M_P^{4n-6}}+...,
\eeq
Obviously, for any realistic value of $m_{3/2}$,
even after tuning the theory to have
$V_Q\lesssim  H_0^2M_P^2$ at its minimum,
one can not keep this $V_Q$ within the range of
$H_0^2M_P^2$ over the Planck scale range of $Q$.

\section{heterotic $M$
or Type I string axion  as quintessence}

In the previous section, we argued that non-metrical quintessence
should corrrespond  to the Goldstone boson of an almost exact 
nonlinear  global symmetry  in order to avoid infinite number
of fine tuning problems associated with the  flat
potential. It is also noted  that explicit breaking of this
global symmetry by quantum gravity effects must  be extremely tiny.
In this section, we discuss the possibility that
string and/or $M$  theory axions correspond to 
such   Goldstone-type quintessence.
There are three different classes of string/$M$ theory vacua
which are particularly interesting in  phenomenological sense: 
perturbative heterotic string vacua \cite{polchinski},
heterotic $M$ theory vacua \cite{horava}, and Type I string vacua  
with $D$-brane configurations \cite{munoz}. 
It is well known that axions in perturbative heterotic string
vacua  receives a large potential energy associated with
the string world sheet instanton effects \cite{worldsheetinstanton}, 
so here we concentrate
on heterotic $M$ theory and Type I string vacua.
We will see that certain combination of the heterotic $M$ theory
or Type I string theory axions can be a plausible candidate
for quintessence if some conditions on the moduli dynamics
are satisfied.

Let us first consider the  heterotic $M$-theory  on a
11-dimensional manifold with boundary which is invariant
under the $Z_2$-parity \cite{horava}:
\beq
C\rightarrow -C,
\quad
x^{11}\rightarrow -x^{11},
\eeq
where $C=C_{ABC}dx^Adx^Bdx^C$ is the 3-form field in the 11-dimensional supergravity.
When compactified to 4-dimensions, axions arise
as the massless modes of $C_{\mu\nu 11}$ and $C_{mn 11}$:
\beq
\epsilon^{\mu\nu\rho\sigma}\partial_{[\nu}C_{\rho\sigma]11}
=\partial^{\mu}\eta_S,
\quad
C_{mn 11}=\sum_i \eta_i(x^{\mu}) \omega^i_{mn},
\eeq
where $\omega^i$ ($i=1$ to $h_{1,1}$) form the basis of the integer $(1,1)$ cohomology
of the internal 6-manifold and  $\mu$, $\nu$ are tangent to 
the noncompact 4-dimensional spacetime.
In 4-dimensional effective supergravity, these axions appear as 
the pseudo-scalar components of chiral multiplets:
\beqar
&&S=(4\pi)^{-2/3}\kappa^{-4/3}V+i\eta_S,
\nonumber \\
&&T_i=(4\pi)^{-1/3}\kappa^{-2/3}\int_{{\cal C}_i}\omega\wedge dx^{11}
+i\eta_i,
\label{st}
\eeqar
where $\kappa^2$ denotes the 11-dimensional gravitational coupling,
$V$ is the volume of the internal 6-manifold with the K\"ahler
two form $\omega$, and
the integral  is over the 11-th segment and   
also over the 2-cycle ${\cal C}_i$ dual to $\omega^i$.
Here the axion components are normalized by the
discrete Peccei-Quinn (PQ) symmetries:
\beq
{\rm Im}(S)\rightarrow {\rm Im}(S)+1,
\quad
{\rm Im}(T_i)\rightarrow {\rm Im}(T_i)+1,
\label{discretepq}
\eeq
which are the parts of discrete modular symmetries.

Holomorphy and the discrete PQ symmetries imply that
in the large ${\rm Re}(S)$ and ${\rm Re}(T_i)$ limits
the gauge kinetic functions can be written as
\beq
4\pi f_a=k_a S+\sum_il_{ai}T_i+
{\cal O}(e^{-2\pi S}, e^{-2\pi T_i}),
\label{gaugekinetic}
\eeq
where $k_a$ and $l_{ai}$ are model-dependent {\it quantized} real constants
and the exponentially suppressed terms  are possibly
due to  the membrane or 5-brane instantons.
For a wide class of compactifed heterotic $M$-theory, 
we have \cite{banks,stieberger}
\beqar
4\pi f_{E_8}&=& S+\sum_i l_iT_i+{\cal O}(e^{-2\pi S}, e^{-2\pi T_i}),
\nonumber \\
4\pi f_{E_8^{\prime}}&=& S-\sum_i l_i T_i
+{\cal O}(e^{-2\pi S}, e^{-2\pi T_i}),
\label{mgaugekinetic}
\eeqar
where $l_iT_i$ corresponds to the one-loop threshold
correction 
in perturbative heterotic string terminology
with the quantized coefficients $l_i$ determined by the instanton numbers
on the hidden wall and also the orbifold twists.

Let $Q$ denote a linear combination of ${\rm Im}(T_i)$
{\it orthogonal} to  the combination $\sum_i l_{i}{\rm Im}(T_i)$
and define the associated nonlinear global symmetry:
\beq
U(1)_Q: \quad Q\rightarrow Q+{\rm constant}.
\label{Q-pqsymmetry}
\eeq
Note that such $Q$  exists always  as long as $h_{1,1}>1$.
In this regard, models of particular interest are the
recently discovered threshold-free models with $l_i=0$
\cite{stieberger} for which
any of ${\rm Im}(T_i)$ can be identified as $Q$.
At any rate,  $Q$ can be a candidate for quintessence if 
$U(1)_Q$ breaking couplings in the effective supergravity
are all suppressed as in Eq.(\ref{bound}). 

It is  easy to see that for $Q$ defined as above
\beq
\frac{\partial f_a}{\partial Q}={\cal O}(e^{-2\pi T}),
\label{smallgauge}
\eeq
over the entire  range of $Q$.
Here we use the unit with $M_P=1$ and
the internal 6-manifold is assumed to be isotropic, so
${\rm Re}(T_i)\ap {\rm Re}(T)$ for all $T_i$.
Holomorphy and discrete PQ symmetries
imply also
\beq
\frac{\partial Y_{ijk}}{\partial Q}={\cal O}(e^{-2\pi T})
\label{smallyukawa}
\eeq
for the Yukawa couplings $Y_{ijk}$
in the superpotential.
Similar estimate applies also to the non-renormalizable
holomorphic couplings in the superpotential.
A non-perturbative  superpotential $W_0$ may be induced
by these holomorphic gauge and superpotential couplings,
e.g. by gaugino condensation.
The $U(1)_Q$ breaking in $W_0$  is  estimated to be
\beq
\frac{\partial W_0}{\partial Q}=
\frac{\partial W_0}{\partial f_a}
\frac{\partial f_a}{\partial Q}+
\frac{\partial W_0}{\partial Y_{ijk}}
\frac{\partial Y_{ijk}}{\partial Q}+...
={\cal O}(e^{-2\pi T}W_0).
\label{smallsuper}
\eeq
We then have 
\beq
\frac{\partial W}{\partial Q}={\cal O}(e^{-2\pi T}W)
\label{smallsuper1}
\eeq
for the full superpotential $W$.

The above estimates of  $U(1)_Q$ breaking couplings
in  $f_a$ and $W$ are made  by  simple macroscopic argument based on
supersymmetry and discrete PQ symmetries. However
one can easily identify its microscopic origin by noting that
$2\pi {\rm Re}(T_i)=(4\pi)^{-1/3}\kappa^{-2/3}\int_{{\cal C}_i}\omega\wedge
dx^{11}$ corresponds to the Euclidean action of
the membrane instanton wrapping ${\cal C}_i$ and stretched along
the 11-th segment \cite{banks}.
When extrapolated to the perturbative heterotic string vacua,
such membrane instanton corresponds to the heterotic string
worldsheet instanton wrapping the same 2-cycle \cite{worldsheetinstanton}.
Explicit computations then show that 
the K\"ahler potential and/or the gauge kinetic functions
are indeed corrected by worldsheet instantons, yielding 
$\delta K={\cal O}(e^{-2\pi T})$ and $\delta f_a={\cal O}(e^{-2\pi T})$
\cite{worldsheetinstanton1}.
These corrections  can be smoothly extrapolated back to
the heterotic $M$-theory vacua \cite{ckm}
and identified as the corrections induced
by stretched membrane instanton.
If a nonperturbative superpotential
$W_0$ is generated by  gaugino condensation, 
$\delta f_a={\cal O}(e^{-2\pi T})$ leaves its trace in $\delta W_0=
{\cal O}(e^{-2\pi T}W_0)$.
We thus conclude that $U(1)_Q$ breaking terms of ${\cal O}(e^{-2\pi T})$
are indeed induced in $K$, $W$ and $f_a$ for generic
heterotic $M$ theory vacua.

The holomorphy  and discrete PQ symmetries (\ref{discretepq})
ensure that $U(1)_Q$ breaking terms  in $f_a$ and $W$,
whatever their microscopic origin is,
are all suppressed by $e^{-2\pi T}$, 
Also  the $U(1)_Q$ breaking terms in $K$ induced by stretched
membrane instanton are  suppressed by $e^{-2\pi T}$,
However there may be unsuppressed  $U(1)_Q$ breaking term in
non-holomorphic $K$, which would arise from 
yet unknown microscopic origin.
Although not definite, it is unlikely to have such 
unsuppressed correction. $U(1)_Q$ is a linear combination
of $U(1)_i: \delta T_i=
ic_i$ ($c_i=$ real constant) which  originate from
the {\it local} transformation of the 11-dimensional three form field: 
$\delta C=
\omega^i\wedge dx^{11}$
with $c_i=\int \delta C=\int \omega^i\wedge dx^{11}$.
This indicates that upon ignoring the effects of boundary 
degrees of freedom, 
the zero momentum mode of ${\rm Im}(T_i)$
couples only to  stretched membrane instantons
wrapping  the 2-cycle ${\cal C}_i$
and are  stretched between the  boundaries.
Since we chose the combination to avoid the breaking by Yang-Mills
instantons on the boundary,
$U(1)_Q$  appears to be  broken only by 
stretched membrane instantons 
whose effects are suppressed by $e^{-2\pi T}$.
It is thus expected that  
\beq
\frac{\partial K}{\partial Q}={\cal O}(e^{-2\pi T})
\label{smallkahler}
\eeq
as in the case of $f_a$ and $W$.

Obviouslly Eqs. (\ref{smallgauge}),
(\ref{smallsuper1}), and (\ref{smallkahler}) show that $U(1)_Q$ 
becomes  an almost exact global symmetry   
in the limit  ${\rm Re}(T)\gg 1$.
From the supergravity potential
\beq
V_{sugra}=e^K\left[K^{I\bar{J}}D_IW(D_JW)^*-3|W|^2\right]
\label{sugrapotential}
\eeq
and also the standard order of magnitude relations
$m_{3/2}\sim W/M_P^2 \sim D_IW/M_P$, 
one easily finds that the axion potential is given by
\beq
V_Q\sim e^{-2\pi{\rm Re}(T)}
m_{3/2}^2M_P^2\cos[2\pi {\rm Im}(T)].
\label{maxionpotential}
\eeq
This axion potential can be identified as the quintessence
potential $V_Q\sim (3\times 10^{-3} {\rm eV})^4$ 
if the modulus vacuum value is
given by
\beq
{\rm Re}(T)
\sim\frac{1}{2\pi}\ln(m_{3/2}^2M_P^2/V_Q)
\sim 32,
\label{modulusvev}
\eeq
where $m_{3/2}\sim 10^2$ GeV are used for numerical estimate.

In the above discussion, gauge kinetic functions are assumed
to be given by (\ref{mgaugekinetic}). In such cases,
$Q$ must be  a linear combination of  ${\rm Im}(T_i)$
to avoid the couplings to the QCD and/or the  hidden gauge anomaly.
Gauge kinetic functions  can be generalized to 
the form of (\ref{gaugekinetic}) with {\it non-universal}
$k_a$, and then $Q$ can include the ${\rm Im}(S)$ component.
At any rate, as long as the quantized coefficients
$k_a$ and $l_{ai}$ are all of order unity,
the domain of moduli space allowing a quintessence axion in heterotic
$M$  theory has
\beq
{\rm Re}(S)={\cal O}\left(\frac{1}{\alpha_{GUT}}\right),
\quad {\rm Re}(T)={\cal O}\left(\frac{1}{\alpha_{GUT}}\right).
\label{valueofst}
\eeq
On this domain, we also have 
\beqar
&& \frac{M_{GUT}}{M_P}\sim [{\rm Re}(S){\rm Re}(T)]^{-1/2}
={\cal O}(\alpha_{GUT}),
\nonumber \\
&& \frac{\kappa^{2/3}}{(\pi\rho)^3}\sim 
{\rm Re}(S)/[{\rm Re}(T)]^3={\cal O}(\alpha^2_{GUT}),
\eeqar
implying that this domain gives  the unification scale $M_{GUT}$
close to the phenomenologically
favored value $3\times 10^{16}$ GeV, and also
the length $\pi\rho$ of the 11-th segment  about one
order of magnitude bigger than the 11-dimensional Planck length
$\kappa^{2/9}$.

Let us now turn to  Type I string axions.
The Type I axions (again normalized by
the discrete PQ symmetries of (\ref{discretepq}))
correspond to the massless
modes of the R-R two form fields $B_{\mu\nu}$ and $B_{mn}$:
\beq
\epsilon^{\mu\nu\rho\sigma}\partial_{\nu}B_{\rho\sigma}
=\partial^{\mu}\eta_S, \quad
B_{mn}=\sum_i\eta_i(x^{\mu})\omega^i_{mn},
\eeq
and they    form 4-dimensional chiral multiplets 
together with the string dilaton
$e^D$ and the internal space volume $V$:
\beqar
&&S=(2\pi)^{-6}\alpha^{\prime -3}e^{-D}V+i\eta_S,
\nonumber \\
&&T_i=(2\pi)^{-2}\alpha^{\prime -1}e^{-D}\int_{{\cal C}_i}\omega
+i\eta_i.
\eeqar
Here we include $D9$ and $D5$ branes in the vaccum configuration,
and consider the 4-dimensional gauge couplings
$\alpha_9$ and $\alpha_{5i}$ defined on
$D9$ branes wrapping the internal 6-manifold and 
$D5$ branes wrapping the 2-cycles ${\cal C}_i$, respectively
\cite{munoz}.
Again holomorphy and discrete PQ symmetries imply that the corresponding
gauge kinetic functions can be written as
(\ref{gaugekinetic}).
A simple leading order calculation gives $\alpha_9=1/{\rm Re}(S)$
and $\alpha_{5i}=1/{\rm Re}(T_i)$, and so \cite{munoz}
\beq
4\pi f_9= S, \quad
4\pi f_{5i}= T_i.
\label{leading}
\eeq
However this leading order result
can receive  perturbative and/or non-perturbative corrections.
Generic perturbative corrections can be expanded
in powers of the string coupling $e^{D}$, the
string inverse tension $\alpha^{\prime}$,
and also the inverse tension $e^D\alpha^{\prime k}$ of $D_{2k-1}$ branes
\cite{polchinski}.
Generically they scale as
\beq
e^{nD}\alpha^{\prime m}
\propto 
[{\rm Re}(S)]^{\frac{n-m}{2}}
[{\rm Re}(T)]^{\frac{m-3n}{2}}.
\eeq
Combined with the leading order result (\ref{leading}) and also
the general form of gauge kinetic function
(\ref{gaugekinetic}) dictated by holomorphy and discrete PQ symmetries,
this scaling behavior implies that
$f_9$ can receive a $T_i$-dependent correction
at order $\alpha^{\prime 2}$, while there is no perturbative correction to $f_{5i}$,
and so
\beqar
4\pi f_9&=& S+l_{i}T_i+{\cal O}(e^{-2\pi S},e^{-2\pi T_i}),
\nonumber \\
4\pi f_{5i}&=& T_i+{\cal O}(e^{-2\pi S}, e^{-2\pi T_i}).
\label{type1gauge}
\eeqar

Similarly to the case of heterotic $M$-theory,
the quintessence axion  $Q$ can arise   as a linear combination of 
${\rm Im}(S)$ and ${\rm Im}(T_i)$, however its explicit form
depends on how the various gauge couplings  are embedded in the model.
Note that for the QCD  and even stronger hidden sector gauge couplings,
both ${\rm Re}(f_a)$ and ${\rm Im}(f_a)$ 
are required to be $Q$-independent as in (\ref{bound}), while for
the weaker gauge interactions
${\rm Im}(f_a)$ are allowed to have a sizable $Q$-dependence.
Here are some  possibilities.
If all of the standard model  gauge couplings and the stronger
hidden sector gauge couplings  are embedded in $\alpha_{5i}$,
${\rm Im}(S)$ can be a quintessence when 
${\rm Re}(S)\sim 32$.
If  those gauge couplings  are embedded in $\alpha_9$,
any of ${\rm Im}(T_i)$ can be a quintessence 
when ${\rm Re}(T_i)\sim 32$. 
If the vacuum does not include any 
$D5$ brane wrapping the particular 2-cycle, e.g. the $i$-th cycle,
the corresponding axion ${\rm Im}(T_i)$ can be a quintessence
independently of the embedding when ${\rm Re}(T_i)\sim 32$.

For the quintessence component $Q$ defined as above,
the $Q$-dependences of $K$, $W$ and $f_a$
are all suppressed by $e^{-2\pi Z}$ ($Z=S$ or $T$) as 
in the case of heterotic $M$-theory.
So the quintessence axion potential is again given by
(\ref{maxionpotential}) where now $T$ is replaced  by $Z$.
Microscopic origin of this Type I axion potential can be easily identified also
by noting  that $2\pi {\rm Re}(S)$ corresponds to 
the Euclidean action of $D5$ brane instanton wrapping the internal
6-manifold and $2\pi {\rm Re}(T_i)$ is of $D1$ string instanton wrapping
the 2-cycle ${\cal C}_i$.
Quintessence axion in Type I
string theory requires also that 
both  ${\rm Re}(S)$ and ${\rm Re}(T)$ are of ${\cal O}(1/\alpha_{GUT})$.
On this domain of moduli space, we  have
\beqar
&&  e^{2D}\sim {\rm Re}(S)/[{\rm Re}(T)]^3={\cal O}(\alpha_{GUT}^2),
\nonumber \\
&& \alpha^{\prime}/V^{1/3}\sim \left[{\rm Re}(T)/
{\rm Re}(S)\right]^{1/2}={\cal O}(1),
\eeqar
and thus a rather weak string coupling and also  a rather
strong  sigma model  coupling.  

So far, we have noted that
a certain combination of axions in  heterotic $M$ theory
or Type I string theory can be a candidate for quintessence
if its modulus partner has a large vacuum value.
For the  axion component normalized as
${\rm Im}(Z)\equiv {\rm Im}(Z)+1$, explicit breaking of the associated  
global symmetry ($U(1)_Q: {\rm Im}(Z)\rightarrow {\rm Im}(Z)+{\rm constant}$)
is suppressed  by $e^{-2\pi Z}$. More concretely,
the effective supergravity
model is described by
\beqar
&& K=\tilde{K}(Z+Z^*)+{\cal O}(e^{-2\pi Z}),
\nonumber \\
&& f_a=\tilde{f}_a+{\cal O}(e^{-2\pi Z}),
\nonumber \\
&& W=[1+{\cal O}(e^{-2\pi Z})]\tilde{W},
\eeqar
where $f_a$  denote the gauge kinetic functions
for the standard model gauge group and also the stronger
hidden sector gauge group, and  $\tilde{f}_a$ and $\tilde{W}$ 
are {\it  $Z$-independent}.
We then have
\beq
{\cal L}_Q=
\frac{1}{2}(\partial_{\mu} Q)^2-
m^4[\cos (Q/v_Q)+1],
\label{axionlagrangian}
\eeq
where
\beq
m^4\sim 
e^{-2\pi {\rm Re}(Z)}m_{3/2}^2M_P^2
\sim (3\times 10^{-3} {\rm eV})^4,
\eeq
for ${\rm Re}(Z)\sim 32$,
and  the canonical quintessence axion $Q$ and its decay constant $v_Q$ 
are given by
\beq
Q=M_P\sqrt{2 K^{\prime\prime}} \, {\rm Im}(Z),
\quad
v_Q=\frac{1}{2\pi}M_P\sqrt{2K^{\prime\prime}},
\label{canonicalq}
\eeq
where $K^{\prime\prime}$ denotes the K\"ahler metric of $Z$:
\beq
K^{\prime\prime}=\frac{\partial^2 K}{\partial Z\partial Z^*}.
\eeq
At leading order approximation in perturbation theory,
we have ${K}=c\ln(Z+Z^*)$ for a constant $c$ of order unity,
and then ${\rm Re}(Z)$ can not be stabilized at the desired
value. Obviously $W$ is almost $Z$-independent
for large  ${\rm Re}(Z)$, so can not provide a stabilizing  
potential of ${\rm Re}(Z)$. Note that if $W$  provides any
sizable potential of ${\rm Re}(Z)$, it means also
a sizable potential of ${\rm Im}(Z)$, which is not allowed
for ${\rm Im}(Z)$ to be a  quintessence.
We thus need  a rather strong $U(1)_Q$-preserving
nonperturbative effects encoded in the K\"ahler potential,
e.g. $K_{np}\sim (Z+Z^*)^{k}e^{-b\sqrt{Z+Z^*}}$,
which would stabilize   ${\rm Re}(Z)$ at
${\rm Re}(Z)\sim 32$ \cite{banks1}. 
As we will discuss in the next section, such strong
correction to $K$ is required also to enlarge  $K^{\prime\prime}$
for cosmological reasons.

\section{Initial condition problem  and  a late time inflation solution:}

In order to suppress $U(1)_Q$  breaking quantum gravity effects,
we needed  a large vacuum value of ${\rm Re}(Z)$.
A large value of ${\rm Re}(Z)$  then implies $K^{\prime\prime}
={\cal O}([{\rm Re}(Z)]^{-2})\ll 1$, and so
$v_Q\ll M_P$. (See Eq. (\ref{canonicalq}).)
Since $v_Q$ determines the coupling  strength of $Q$, while
$M_P$ determines the gravitational coupling strength,
$Q$ responds to its
potential energy more sensitively  than the expanding universe  does.
As a result, generic initial values  of $Q$ and $\dot{Q}$
give  a rapidly rolling $Q$ and thus   positive
pressure at present. To avoid this,
we   need   a fine tuning  of  initial condition.
So the attempt to  avoid the fine tunings of parameters for the flat potential
leads to a new fine tuning problem.
In this section, we propose a late  time inflation scenario based on
the modular  and CP invariance which would solve this initial condition
problem.

Let us first consider what kind of initial conditions we need
to have.
When applied for (\ref{axionlagrangian}) and (\ref{canonicalq}),
the slow roll condition (\ref{slowroll1}) 
leads to
\bea
&& |2\pi {\rm Im}(Z)|\ler 
\sqrt{2 K^{\prime\prime}}/2\pi=
{\cal O}\left(\frac{1}{2\pi {\rm Re}(Z)}\right), 
\nonumber \\
&& |2\pi{\rm Im}(\dot{Z})|\ler
2\pi H_0/\sqrt{2 K^{\prime\prime}}={\cal O}\left(\frac{2\pi H_0}{
{\rm Re}(Z)}\right),
\label{slowrollcondition}
\eea
implying that ${\rm Im}(Z)$ should be
at near the top of its potential.
The main difficulty of this condition is that 
\beq
m_Q\equiv \left|\frac{\partial^2  V_Q}{\partial Q^2}\right|^{1/2}_{Q=0}
=\frac{m^2}{v_Q}\approx \frac{\sqrt{3}\pi}{\sqrt{K^{\prime\prime}}}
H_0\gg H_0,
\label{axionmass}
\eeq
and thus  a small ${\rm Im}(Z)$  is  unstable 
against the cosmlogical evolution during the period
with an expansion
rate $H\lesssim m_Q$.
The resulting instability factor is given by
$e^{\gamma m_Q/H_0}$ where $\gamma$ is a constant
of order unity whose precise value  depends upon the initial conditions.
A detailed numerical study of the cosmological evolution gives
$\gamma\approx 0.5$ for  wide range of relevant initial conditions
\cite{hclee}.
Thus the slow-roll condition (\ref{slowrollcondition}) requires
eventually the following initial conditions 
\beqar
&&
|2\pi {\rm Im}(Z)|_{in}\lesssim e^{-\gamma m_Q/H_0}
\sim e^{-3/\sqrt{K^{\prime\prime}}},
\nonumber \\
&&
|2\pi {\rm Im}(\dot{Z})|_{in}\lesssim  H 
e^{-\gamma m_Q/H_0}\sim H e^{-3/\sqrt{K^{\prime\prime}}},
\label{initialcondition}
\eeqar
for the period with $H\gg m_Q$.

Eq. (\ref{initialcondition}) shows that
the degree of required fine tuning is quite sensitive to the value
of the K\"ahler metric $K^{\prime\prime}$.
At leading order approximation in string or $M$ theory,
we have \cite{polchinski}
$K\ap -c\ln(Z+Z^*)$ where $c$ is a constant of order unity,
so $K^{\prime\prime}={\cal O}(10^{-3})$
for ${\rm Re}(Z)\sim 30$.
As we will see, the degree of  fine tuning 
for  this value of $K^{\prime\prime}$
is too  severe to be accommodated,
so we need a mechanism to enlarge the value of  $K^{\prime\prime}$.
It is expected  that $K^{\prime\prime}$
can be  enlarged  by $U(1)_Q$-preserving
nonperturbative effects which would be responsible
for stabilizing ${\rm Re}(Z)$ \cite{banks1}. 
However it is hard to imagine that 
$K^{\prime\prime}$ becomes  of order unity, which would allow to
avoid the fine tuning of initial condition without any further
mechanism.
In the following,  we discuss a late time inflation scenario
which would resolve the fine tuning problem of initial
conditions for a reasonably enlarged    value of $K^{\prime\prime}$.

The gauge symmetries of string or $M$-theory include discrete modular group
\cite{polchinski}
under which $Z$ and 
other generic moduli $\phi$ transform as
\beq
{\rm Re}(Z)\rightarrow
\frac{1}{{\rm Re}(Z)},
\quad
{\rm Im}(Z)\rightarrow {\rm Im}(Z)+1,
\quad
\phi\rightarrow\phi^{\prime},
\eeq
and also CP \cite{gaugecp} under which
\beq
Z\rightarrow Z^*,
\quad
\phi\rightarrow \phi^*.
\eeq
Here we take the simplest  form of
the $Z$-duality ($Z=S$ or $T$),
i.e. ${\rm Re}(Z)\rightarrow 1/{\rm Re}(Z)$
with the self-dual value ${\rm Re}(Z)=1$,
however our discussion is valid for
other forms of the $Z$-duality transformation
as long as the self-dual value is of order unity.

The modular and CP invariance  ensure 
that  the invariant points,
\beq
{\rm Re}(Z)=1,
\quad
{\rm Im}(Z)=0 \, \, \, {\rm or} \, \, \, \frac{1}{2},
\quad
\phi=\phi^* \, \, \, {\rm or} \, \, \, \phi^{\prime *},
\eeq
correspond to the stationary points of
the effective action \cite{nir}.
It is then quite possible that
the modular invariant point $Z=1$ is  a (local) minimum
of the effective  potential {\it during  the inflationary period}
if the inflaton is a modular invariant field.
This (local) minimum may become unstable if the inflaton field
takes  the present value, making ${\rm Re}(Z)$
rolls toward the present minimum at ${\rm Re}(Z)\sim 32$ after inflation.  
At  any rate, during the inflationary phase, 
${\rm Re}(Z)\sim 1$ and then
all the moduli masses including that of ${\rm Im}(Z)$
have the same order of magnitude:
\beq
m_{{\rm Re}(Z)}\sim
m_{{\rm Im}(Z)}\sim m_{\phi}\sim  H_{inf},
\eeq
where $H_{inf}$ denotes the expansion rate during inflation.
Note that the axion mass $m_{{\rm Im}(Z)}$ is unsuppressed
for ${\rm Re}(Z)\sim 1$.
Since ${\rm Re}(Z)$ was far  away from the present value,
it  is expected that the inflationary potential
is at least of ${\cal O}(m_{3/2}^2M_P^2)$,
and so $H_{inf}$ is at least of ${\cal O}(m_{3/2})$.
To avoid a too large quantum fluctuation during this  inflation,
we take the minimal value of $H_{inf}$, so
\beq
H_{inf}={\cal O}(m_{3/2}).
\eeq

About the location of the minimum  in the axion direction,
we have just two possibilities
if CP is {\it not} spontaneously broken in the moduli sector.
One of the two CP invariant points,  ${\rm Im}(Z)=0$  and
$1/2$, is the minimum, while
the other  is the maximum.
{\it Our key assumption is that  
${\rm Im}(Z)=0$ was the minimum for the
inflationary modulus value ${\rm Re}(Z)=1$,
however it becomes the maximum for
the present modulus value ${\rm Re}(Z)\sim 32$.}
Note that  the coefficient of the cosine potential of ${\rm Im}(Z)$
is a function of ${\rm Re}(Z)$, and so its sign
can be changed  when ${\rm Re}(Z)$ varies from 
the inflationary value to the present value.

Given the features of the moduli potential discussed above,
during the inflationary period all  moduli are settled down 
near at the modular and CP-invariant 
local minimum with ${\rm Re}(Z)\approx 1$, 
${\rm Im}(Z)\approx 0$,
and $\phi\approx \phi^*$.
In particular, just after inflation, we have
\cite{randall}
\bea
&& |2\pi {\rm Im}(Z)|_{in}= {\cal O}(e^{-3N_e/2})+
{\cal O}(H_{inf}/M_P),
\nonumber \\
&& |2\pi {\rm Im}(\dot{Z})|_{in}=
{\cal O}( e^{-3N_e/2} H_{inf})+
{\cal O}(H_{inf}^2/M_P),
\label{inflationvalue}
\eea
where $N_e$ denotes  the number of efoldings.
Here  the exponential suppression is due to the classical evolution
toward the minimum of the inflationary potential,
while the second terms represent
the quantum fluctuations.

After this late inflation,
${\rm Re}(Z)$ rolls toward the present minimum at ${\rm Re}(Z)\sim 32$.
In this period, the  coefficient of the axion potential
changes its sign, and thus
${\rm Im}(Z)=0$ which was  the minimum at the
inflationary phase becomes the maximum
of the present axion potential.
The small value of ${\rm Im}(Z)$, i.e. (\ref{inflationvalue}),
which was set during the inflation
becomes unstable if $H$ starts to be smaller than $m_Q$.
However  if the inflationary values  
 (\ref{inflationvalue})
satisfy the condition (\ref{initialcondition}),
the present value of the quintessence axion satisfies
(\ref{slowrollcondition}), and thus provides
an accelerating universe at  present.
This requires  a  large number of efolding
\beq
N_e\gtrsim \frac{2\gamma m_Q}{3H_0}\approx
\frac{2}{\sqrt{K^{\prime\prime}}},
\label{efolding}
\eeq
and also  a small quantum fluctuation
\beq
\delta Q\sim \frac{H_{inf}}{2\pi} \lesssim 
M_Pe^{-3/\sqrt{K^{\prime\prime}}}.
\label{quantum} 
\eeq
However the number of efolding can not be arbitrarily large.
For a relatively late inflation with $H_{inf}\sim m_{3/2}$,
it is extremely diffcult to generate the observed density
fluctuation $\delta \rho/\rho\sim 10^{-5}$.
It is thus reasonable to assume that density fluctuations were
created before the late inflation. In this case, the late inflation
is required not to destroy the  pre-exisiting density fluctuations.
This consideration gives an upper bound on $N_e$ \cite{randall},
yielding 
\beq
N_e\lesssim 25\sim 30
\label{structure}
\eeq
for a weak scale $m_{3/2}$ and the
reheat temperature range $T_r=10^5\sim 10^{-2}$ GeV.
Combining (\ref{efolding}), (\ref{quantum}) and
(\ref{structure})  together, we find that our late time inflation
scenario  can successfully generate the required 
initial condition (\ref{initialcondition}) if
the K\"ahler metric is enlarged to be 
\beq
K^{\prime\prime}\approx  8\times 10^{-3}.
\eeq
It is not unreasonable to expect that
the nonperturbative terms  in the K\"ahler potential
stabilizing ${\rm Re}(Z)$ provide the necessary enlargement
of the K\"ahler metric,
particularly when
those terms  involve a large power of $(Z+Z^*)$, 
e.g. $K_{np}= h(Z+Z^*)^ke^{-b\sqrt{Z+Z^*}}={\cal O}(1)$
for ${\rm Re}(Z)\sim 32$ with $k\gtrsim 6$,

\medskip


\end{document}